\documentclass[fleqn,twoside]{article}
\usepackage{espcrc2}

\usepackage{graphicx}
\usepackage[figuresright]{rotating}

\newcommand{\AmS}{{\protect\the\textfont2
  A\kern-.1667em\lower.5ex\hbox{M}\kern-.125emS}}

\title{Flavor and CP Violation Induced by Atmospheric Neutrino Mixing}

\author{Junji Hisano \address[MCSD]{ICRR, University of Tokyo,  5-1-5 Kashiwa-no-Ha
Kashiwa City, 277-8582, Japan}%
}
       
\begin{document}

\begin{abstract}
Neutrino oscillation experiments suggest existence of new
flavor-violating interactions in high energy scale. It may be
possible to probe them by the flavor- and CP-violating processes in
leptons and hadrons in the supersymmetric (SUSY) models. In this
article the third-generation flavor violation is reviewed in the SUSY
seesaw model and SUSY SU(5) GUT with the right-handed neutrinos.
\vspace{1pc}
\end{abstract}

\maketitle

\section{Introduction}
\label{sec:tau1}

After discovery of the atmospheric neutrino oscillation by the
superKamiokande experiment in 98' \cite{skatm}, it is found that the
lepton sector has much different flavor structure from the quark
sector \cite{Ahmad:2002jz}\cite{kamland}.  The seesaw mechanism is the
most promising model to explain the tiny neutrino masses and the
neutrino oscillation \cite{seesaw}.  However, the observed mixing
angles between the first and second and between second and third
generations are almost maximal, and those are different from the naive
expectation in the extension of the seesaw mechanism to the grand
unified theories (GUTs).  Various attempts to understand those
mismatches between quark and lepton mixing angles have been made so
far.

We may find new clues to understand the origin of the neutrino masses
in the future precision experiments of hadrons and leptons.  In the
supersymmetric (SUSY) models, the flavor- and CP-violating phenomena in
hadron and lepton physics are windows to probe flavor structure in the
high energy physics, such as the seesaw mechanism and the GUTs.  The
imprints may be generated in the SUSY-breaking slepton or squark mass
terms if the SUSY-breaking terms are generated at the higher than the
energy scale. In this case the SUSY-breaking terms induce the sizable
effect to the flavor- and CP-violating processes in hadrons and
leptons, since they are suppressed by only the SUSY-breaking scale,
not the energy scale responsible to the flavor violation such as the
right-handed neutrino mass or GUT scale \cite{Hall:1985dx}.

In the SUSY seesaw mechanism the radiative correction by the neutrino
Yukawa coupling induces the left-handed slepton mixing even if the
soft SUSY-breaking terms are universal at the GUT or Planck
scale. This lepton-flavor violation (LFV) leads to the radiative LFV
decay of tau and muon \cite{bm}. If the SUSY seesaw model is extended
to the SUSY GUTs, the right-handed down-type squarks also have the
flavor-violating SUSY-breaking mass terms since they are embedded in
common SU(5) multiplets with the left-handed leptons
\cite{Barbieri:1995rs}\cite{moroi}. This may lead to new flavor- and
CP-violating source.

The Belle and BaBar experiments in the KEK and SLAC $B$ factories,
give new information about the flavor violation of the third
generations. The CKM dominance in the flavor- and CP-violating hadron
phenomena is almost confirmed by the experiments.  However, the Belle
experiment reports that the CP asymmetry in $B_d\rightarrow
\phi K_s$ is $3.5 \sigma$ deviated from the Standard-Model (SM)
prediction \cite{Abe:2003yt}. At present the BaBar experiment does not
observe such a large deviation, and the combined result is not yet
significant \cite{babar}. However, the Belle's result might be a
signature of the new physics. In addition to them, the experiments are
improving bounds on the LFV tau decay modes by about one order of
magnitude  \cite{inami}. Now the physics in high luminosity $B$
factories, whose integrate luminosities may reach to the order of
$ab^{-1}$, is discussed. If these super $B$ factories are constructed,
we may get new information about the origin of the neutrino masses.

In this article, we review the third generation flavor violation in
leptons and hadrons by the neutrino Yukawa interaction in
the SUSY seesaw mechanism and the extension to SUSY GUTs, which will
be probed in the super $B$ factories. The atmospheric neutrino
experiments suggest the existence of large flavor violation between
the second and the third generations. The LFV tau decay is a good
probe to the neutrino Yukawa coupling responsible to the atmospheric
neutrino oscillation.

The SUSY GUTs may have rich flavor physics in the hadron and lepton
sectors as mentioned above. However, since the flavor- and
CP-violating phenomena are indirect probes to the new physics, it is
important to take correlation and consistency among various
processes. It is argued that the right-handed sbottom and strange
mixing may induce the sizable deviation for $B_d\rightarrow\phi K_s$
from the SM prediction in this model \cite{moroi}.  On the other hand,
it is pointed out that the deviation is strongly correlated with the
chromoelectric dipole moment (CEDM) of the strange quark, and the size
of the deviation is limited by the experimental bound on the Mercury
EDM \cite{Hisano:2003iw}.

In next section we review the flavor structure in the minimal SUSY
seesaw model, and the prediction for the LFV tau decay modes and the
sensitivities in the future experiments are discussed in Section 3. We
review the flavor structure in the SUSY SU(5) GUT with the
right-handed neutrinos in Section 4, and the EDMs in the SUSY GUTs is
discussed in Section 5. We show the correlation between the EDM of
Mercury and the CP asymmetry in $B_d\rightarrow \phi K_s$
there. Section 6 is devoted to summary.
 
\section{Minimal SUSY Seesaw Model}

The seesaw mechanism is the most fascinating model to explain the
small neutrino masses in a natural and economical way. In this
mechanism the superheavy right-handed neutrinos ${\overline{N}}$ are
introduced, and thus, the supersymmetry is required to stabilize the
hierarchical structure.  In this section we review the flavor structure
in the SUSY-breaking terms in the minimal SUSY seesaw model.

In the minimal SUSY seesaw model only three additional heavy
singlet neutrino superfields are introduced. The relevant leptonic
part of its superpotential is
\begin{eqnarray}
W_{\rm seesaw} & =& f^\nu_{ij} L_i \overline{N}_j  \overline{H}_f
  +  f^l_{ij} \overline{E}_i  L_j H_f 
\nonumber\\
&&+ \frac{1}{2}{M}_{ij} {\overline{N}}_i \overline{N}_j 
\label{MseesawM}
\end{eqnarray}
where the indexes $i,j$ run over three generations and ${M}_{ij}$
is the heavy singlet neutrino mass matrix.  In addition to the three
charged lepton masses, this superpotential has eighteen physical
parameters, including six real mixing angles and six CP-violating
phases, because the Yukawa coupling and the Majorana mass matrices
are given after removing unphysical phases as 
\begin{eqnarray}
f^l_{ij} &=& f_{l_i} \delta_{ij},\\  
f^\nu_{ij} &=& X^\star_{ik} f_{\nu_k} 
{\rm e}^{-i \varphi_{\nu_k}} W^\star_{kj} 
{\rm e}^{-i \overline{\varphi}_{\nu_k}} ,
\\
M_{ij} &=&\delta_{ij} M_{N_k} ,  
\end{eqnarray}
Here, $\sum_i \varphi_{\nu_i}=0$ and  
$\sum_i \overline{\varphi}_{\nu_i}= 0$, and $W$ and $X$ are 
unitary matrices with one phase. Nine parameters associated with the
heavy-neutrino sector cannot be measured in a direct way. The
exception is the baryon number in the universe if leptogenesis is
right
\cite{Fukugita:1990gb}.

At low energies the effective theory after integrating out the right-handed
neutrinos is given by the effective superpotential
\begin{eqnarray}
W_{\rm eff} &=&  f_{l_{i}} \overline{E}_i  L_i H_f 
\nonumber\\
&&  + \frac{1}{2 v^2 \sin^2\beta} ({m_\nu})_{ij} (L_i \overline{H}_f)(L_j \overline{H}_f) \,,
\label{weff}
\end{eqnarray}
where we work in a basis in which the charged lepton Yukawa couplings
are diagonal. The second term in (\ref{weff}) leads to the light
neutrino masses and mixings. The explicit form of the small neutrino
mass matrix $({m_\nu})$ is given by
\begin{eqnarray}
({m_\nu})_{ij} &=&
\sum_k \frac{f^\nu_{ik}f^\nu_{jk}}
            {{M}_{N_k}} v^2 \sin^2\beta \,.
\label{lightMnu}
\end{eqnarray}
The light neutrino mass matrix $({m_\nu})$ is
symmetric, with nine parameters, including three real mixing angles
and three CP-violating phases. It can be diagonalized by a unitary matrix
$Z$ as
\begin{eqnarray}
Z^T {m}_\nu Z &=& {m}^D_\nu\,.
\end{eqnarray}
By redefinition of fields one can rewrite $Z \equiv U P,$ where
$P \equiv {\rm diag}(e^{i\phi_1}, e^{i\phi_2}, 1 )$ and $U$ is the MNS 
matrix, with the three real mixing angles and the remaining CP-violating 
phase.

If the SUSY-breaking terms are generated above the right-handed
neutrino mass scale, the renormalization effects may induce sizable
LFV slepton mass terms, which leads to the LFV charged lepton decay.
If the SUSY-breaking parameters at the GUT scale or the Planck scale
are universal, off-diagonal components in the left-handed slepton mass
matrix $(m^2_{L})$ and the trilinear slepton coupling $(A_e)$ take the
approximate forms 
\begin{eqnarray}
(\delta m_{{L}}^2)_{ij}&\simeq&
-\frac{1}{8\pi^2}(3m_0^2+A_0^2) H_{ij} \,,
\nonumber\\
(\delta A_e)_{ij} &\simeq&
-\frac{1}{8\pi^2} A_0 f_{e_i} H_{ij} \,,
\label{leading}
\end{eqnarray}
where $ i\ne j$, and the off-diagonal components of the right-handed
slepton mass matrix are suppressed. Here, the Hermitian matrix $H$, whose
diagonal terms are real and positive, is defined in terms of
$f_\nu$ and the heavy neutrino masses ${M}_{N_k}$ by
\begin{eqnarray}
H_{ij}
&=&
\sum_k 
{f^{\nu\star}_{ik}}
{f^{\nu}_{jk}}
\log\frac{M_{GUT}}{{M}_{N_k}}.
\label{hmatrix}
\end{eqnarray}
Here we take $M_{GUT}$ the GUT scale for simplicity. In
Eq.~(\ref{leading}) the parameters $m_0$ and $A_0$ are the universal
scalar mass and trilinear coupling at the GUT scale.  We ignore terms
of higher order in $f_l$, assuming that $\tan\beta$ is not extremely
large.  

The Hermitian matrix $H$ has nine parameters including three phases,
which are clearly independent of the parameters in $({m}_\nu)$. Thus
two matrices $({m}_\nu)$ and $H$ together provide the required
eighteen parameters, including six CP-violating phases, by which we
can parameterize the minimal SUSY seesaw model \cite{Ellis:2002fe}.

The off-diagonal terms, $H_{ij}(i\ne j)$, are related to the LFV
$l_i-l_j$ transition, and they are related to the LFV charged lepton
decay.  On the other hand, our abilities to measure three phases in
the Hermitian matrix $H$ are limited, in addition to the Majorana
phases $e^{i\phi_1}$ and $e^{i\phi_2}$. Only a phase in $H$ might be
determined by T-odd asymmetries in $\tau\rightarrow3 l$ or
$\mu\rightarrow3 e$ \cite{Ellis:2001xt}, since they are proportional
to a Jarskog invariant obtainable from $H$,
\begin{eqnarray}
J&=&{\rm Im}[H_{12} H_{23} H_{31}]\,.
\end{eqnarray}
In order to determine other two phases in $H$, the asymmetries, which
come from interference between phases in $H$ and $({m}_\nu)$, have to
be measured. A possibility to determine them might be the EDMs
of the charged leptons \cite{Ellis:2001yz}. The threshold correction
due to non-degeneracy of the right-handed neutrino masses might
enhance the EDMs of the charged leptons. They depend on all of the
phases in $({m}_\nu)$ and $H$ in non-trivial ways.

\section{LFV Tau Decay in Minimal SUSY Seesaw Model}

As explained in the previous section we show that the LFV charged
lepton decay gives information about the minimal SUSY seesaw model,
which is independent of the neutrino oscillation experiments. In this
section we demonstrate it by considering the LFV tau decay.

First, we review the LFV charged lepton decay in the SUSY SM. 
In the SUSY models, the LFV processes of the charged leptons are
radiative one due to the $R$ parity. Thus, the largest LFV tau decay
processes are $\tau\rightarrow\mu \gamma$ or $\tau\rightarrow e
\gamma$. Other processes are suppressed by order of $\alpha$. 
They are discussed later. The effective operators relevant to
$l\rightarrow l^{\prime}\gamma$ are flavor-violating dipole moment
operators,
\begin{eqnarray}
\lefteqn{  {H_{\rm eff}}=
  \sum_{l > l^\prime}\frac{4G_F}{\sqrt{2}}m_l
  \left[
    A^{l l^\prime}_R \bar{l}\sigma^{\mu\nu} P_R l^\prime
    +
    A^{l l^\prime}_L \bar{l}\sigma^{\mu\nu} P_L l^\prime
  \right] }
\nonumber\\
\lefteqn{+h.c.,}
  \label{eff_op_lfv}
\end{eqnarray}
where $P_{L/R}=(1\mp\gamma_5)/2$, and the branching ratios
are given as 
\begin{eqnarray}
  Br(l\rightarrow l^{\prime} \gamma) &=& 
  384 \pi^2 (|A^{l l^\prime}_R|^2+|A^{l l^\prime}_L|^2)~
\nonumber\\
&&\times  Br(l\rightarrow l^{\prime} \nu_l \bar{\nu}_{l^\prime}).
\end{eqnarray}
Here, 
$Br(\tau\rightarrow\mu(e)\nu_\tau\bar{\nu}_{\mu(e)})
\simeq 0.17$ and 
$Br(\mu\rightarrow e\nu_\mu\bar{\nu}_e)= 1$.
The coefficients in Eq.~(\ref{eff_op_lfv}) are approximately given
as 
\begin{eqnarray}
  A^{\tau l^\prime}_R
  &=&
  \frac{\sqrt{2}e}{4 G_F} 
  \frac{\alpha_Y}{4\pi} \frac{\tan\beta}{m_{SUSY}^2}
  \left[
    -\frac{1}{120} \delta^R_{\tau l^\prime}
  \right],
  \\
  A^{\tau l^\prime}_L
  &=&
  \frac{\sqrt{2}e}{4 G_F} 
  \frac{\alpha_2}{4\pi} \frac{\tan\beta}{m_{SUSY}^2}
  \left[
    (\frac{1}{30}+\frac{t_W^2}{24}) \delta^L_{\tau l^\prime}
  \right],
  \\
  A^{\mu e}_R
  &=&
  \frac{\sqrt{2}e}{4 G_F} 
  \frac{\alpha_Y}{4\pi} \frac{\tan\beta}{m_{SUSY}^2}
  \left[
    -\frac{1}{120} \delta^R_{\mu e} 
\right.
\nonumber\\
&&
\left.
    +\frac{1}{120} \delta^R_{\mu \tau} \delta^R_{\tau e}
    -\frac{1}{60} \frac{m_\tau}{m_\mu} \delta^L_{\mu \tau}\delta^R_{\tau e}
  \right],
  \\
  A^{\mu e}_L
  &=&
  \frac{\sqrt{2}e}{4 G_F} 
  \frac{\alpha_2}{4\pi} \frac{\tan\beta}{m_{SUSY}^2}
  \left[
    (\frac{1}{30}+\frac{t_W^2}{24}) \delta^L_{\mu e}
\right.
\nonumber\\
&&
    -(\frac{1}{80}+\frac{7t_W^2}{240}) \delta^L_{\mu \tau} \delta^L_{\tau e}
  \nonumber\\
  &&
  \left.
    -\frac{t^2_W}{60} \frac{m_\tau}{m_\mu} \delta^R_{\mu \tau}\delta^L_{\tau e}
  \right],
\end{eqnarray}
assuming for simplicity that all SUSY particle masses are
the same as $m_{SUSY}$ and $\tan\beta\gg 1$. 
Here, $t_W\equiv \tan\theta_W$, where $\theta_W$ is the
Weinberg angle, and the mass insertion parameters are given
as 
\begin{eqnarray}
  \delta^R_{ij}
  = \left(\frac{(m_{\overline{E}}^2)_{ij}}{m_{SUSY}^2}\right),
&&
  \delta^L_{ij}
  = \left(\frac{(m_{L}^2)_{ij}}{m_{SUSY}^2}\right),
\end{eqnarray}
where $(m_{\overline{E}}^2)$ is the right-handed slepton mass matrix.
When the slepton mass matrices are non-vanishing for both $(1,3)$ and $(2,3)$
 components, $\mu\rightarrow e\gamma$ is generated via
stau exchange. Especially, if both the left-handed and
right-handed mixing are sizable, the branching ratio is enhanced by
$(m_\tau/m_\mu)^2$ compared with a case that only left-handed or
right-handed mixing angles are non-vanishing. The off-diagonal
components in $(A_{e})_{ij}$ are sub-dominant in these processes since
the contribution is not proportional to $\tan\beta$.

We list constraints on $\delta^R_{ij}$ and
$\delta^L_{ij}$ from current experimental bounds on
$Br(\tau\rightarrow\mu(e)\gamma)$, which are derived by the
Belle experiment \cite{inami}, and $Br(\mu\rightarrow e\gamma)$ in
Table~\ref{delta_LFV1}.  In this table, we take $\tan\beta=10$ and
$m_{SUSY}=100$GeV and $300$GeV.  The constraints from $\mu\rightarrow
e\gamma$ on the slepton mixings are quite stringent.  On the other
hand, the current bounds on the LFV tau  decay modes give
sizable constraints on $|\delta^L_{\tau\mu}|$ and $|\delta^L_{\tau
e}|$, independently. Furthermore, while the current constraint on
$|\delta^L_{\mu\tau}\delta^L_{\tau e}|$ from the LFV tau decay
is weaker than that from the LFV muon decay, the improvement of the
LFV tau decay modes by about an order of magnitude will give a
competitive  bound $|\delta^L_{\mu\tau}\delta^L_{\tau e}|$ to that from
the LFV muon decay. 

\begin{table*}[htbp]
  \begin{center}
    \begin{tabular}{|c|c|c|c|c|c|}
      \hline
      $m_{SUSY}$ &
      $|\delta^L_{\tau \mu}|$ & 
      $|\delta^L_{\tau e}|$ & 
      $|\delta^L_{\mu e}|$ & 
      $|\delta^L_{\mu\tau }\delta^L_{\tau e}|$ & 
      $|\delta^R_{\mu\tau}\delta^L_{\tau e}|$
      \\
      \hline
      $100$GeV&	
      $2\times 10^{-2}$& 
      $2\times 10^{-2}$& 
      $4\times 10^{-5}$&
      $1\times 10^{-4}$&
      $2\times 10^{-5}$
      \\
      & & & &
      $(3\times 10^{-4})$ &
      $(7\times 10^{-3})$ 
      \\
      \hline
      $300$GeV&	
      $2\times 10^{-1}$& 
      $2\times 10^{-1}$& 
      $4\times 10^{-4}$&
      $9\times 10^{-4}$&
      $2\times 10^{-4}$
      \\
      & & & &
      $(3\times 10^{-2})$ &
      $(6\times 10^{-1})$ 
      \\
      \hline\hline
      ${m_{SUSY}}$&	
      $|\delta^R_{\tau \mu}|$ &
      $|\delta^R_{\tau e}|$ &
      $|\delta^R_{\mu e}|$ &
      $|\delta^R_{\mu\tau }\delta^R_{\tau e}|$ &
      $|\delta^L_{\mu\tau }\delta^R_{\tau e}|$ 
      \\
      \hline
      100GeV&
      $3\times 10^{-1}$ & 
      $3\times 10^{-1}$ &
      $9\times 10^{-4}$ &
      $9\times 10^{-4}$ &
      $2\times 10^{-5}$ \\
      & & & & 
      $(1\times10^{-1})$ & 
      $(7\times 10^{-3})$ \\
      \hline
      300GeV&
      $3$ & 
      $3$ &
      $8\times 10^{-3}$ &
      $8\times 10^{-3}$ &
      $2\times 10^{-4}$ \\
      & & & & 
      $(9)$ & 
      $(6\times 10^{-1})$ \\
      \hline
    \end{tabular}
  \end{center}
  \caption{
    Constraints on $\delta^R_{ll^\prime}$ and 
    $\delta^L_{ll^\prime}$ from current experimental bounds
    on $Br(l\rightarrow l^{\prime}\gamma)$.  Here, we use
    the result for the LFV tau decay search by the Belle experiment \cite{inami}.
    We take $\tan\beta=10$ and $m_{SUSY}=100$GeV and  $300$GeV. 
    The numbers in parentheses are derived from constraints
    on $|\delta^{(L/R)}_{\tau\mu}|$ and 
    $|\delta^{(L/R)}_{\tau e}|$.
  } \label{delta_LFV1}
\end{table*}

Now we present sensitivity of the future experiments for the LFV tau
decay in the minimal SUSY seesaw model.  We take 
two different limits of the parameter matrix $H$ given in
Eq.~(\ref{hmatrix}), of the form \cite{Ellis:2002fe}
\begin{eqnarray}
H_1=\left(\begin{array}{ccc}
a & 0 & 0 \\
0 & b  & d  \\
0 &  d^\dagger & c
\end{array} \right) \, ,
\label{H1}
\end{eqnarray}
and
\begin{eqnarray}
H_2=\left(\begin{array}{ccc}
a & 0 & d \\
0 & b  & 0  \\
d^\dagger &  0 & c
\end{array} \right) \, ,
\label{H2}
\end{eqnarray}
where $a,b,c$ are real and positive, and $d$ is a complex number.  
The non-vanishing $(2,3)$ component in $H_1$ leads to $\tau\rightarrow
\mu \gamma$ while the $(1,3)$ component in $H_2$ does to 
$\tau\rightarrow e \gamma$.

In the above Ansatz, we take $H_{12}=0$ and $H_{13}H_{32}=0$ because
these conditions suppress $Br(\mu\to e\gamma)$.  It is found from the
numerical calculation that $Br(\mu\to e\gamma)$ is suppressed in a
broad range of parameters with the chosen forms $H_1$ and $H_2$. From
a viewpoint of the model-building, the matrix $H_1$ is favored since
it is easier to explain the large mixing angles observed in the
atmospheric and solar neutrino oscillation experiments by structure of
the Yukawa coupling $f_{\nu}$. If we adopt $H_2$, we might have to
require some conspiracy between $f_{\nu}$ and ${M_N}$. However, from a
viewpoint of the bottom-up approach, we can always find parameters
consistent with the observed neutrino mixing angles for both $H_1$ and
$H_2$, as explained in the previous section.

\begin{figure}[htb]
\begin{center}
\includegraphics[width=15pc]{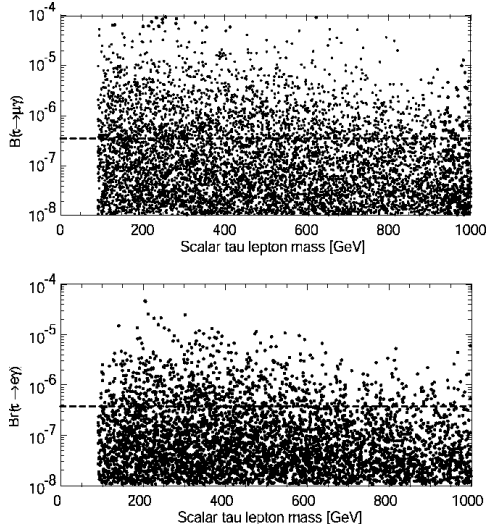} 
\caption{$Br(\tau\rightarrow\mu\gamma)$ for $H_1$ and 
$Br(\tau\rightarrow e\gamma)$ for $H_2$. The input parameters are
given in text.}
\label{fig:lfvtau}
\end{center}
\end{figure}

In Fig.~\ref{fig:lfvtau} we show $Br(\tau\rightarrow\mu\gamma)$ for
the ansatz $H_1$ and $Br(\tau\rightarrow e\gamma)$ for $H_2$ as
functions of the lightest stau mass. We take the SU(2) gaugino mass to
be 200~GeV, $A_0=0$, $\mu>0$, and $\tan\beta=30$ for the SUSY-breaking
parameters in the SUSY SM. We sample the parameters in $H_1$ or $H_2$
randomly in the range $10^{-2}<a,b,c,|d|<10$, with distributions that
are flat on a logarithmic scale. Also, we require the Yukawa
coupling-squared to be smaller than $4 \pi$, so that $f_\nu$ remains
perturbative up to $M_{G}$.

In order to fix the SUSY seesaw model, we take the light neutrino
parameters as $\Delta m^2_{32}=3\times 10^{-3}$ eV$^2,$ $\Delta
m^2_{21}=4.5\times 10^{-5}$ eV$^2,$ $\tan^2\theta_{23}=1$,
$\tan^2\theta_{12}=0.4$, $\sin\theta_{13}=0.1$ and $\delta=\pi/2$. The
Majorana phases $e^{i\phi_1}$ and $e^{i\phi_2}$ are taken randomly.
In Fig.~\ref{fig:lfvtau} we assume the normal hierarchy for the light
neutrino mass spectrum, however, it is found as expected that the
branching ratios are insensitive to  the structure of the light neutrino
mass matrix.

Current experimental bounds for branching ratios of the LFV tau decay modes
are  derived in the Belle experiment, and $ Br(\tau\rightarrow\mu(e)
\gamma) <3.2(3.6)\times 10^{-7}$ \cite{inami}. These results already exclude
a fraction of the parameter space of the minimal SUSY seesaw model.
These bounds may be improved to about $10^{-8}$ in the super $B$ factories
if the signals are not observed
\cite{inami}.\footnote{
If sleptons are found in the future collider experiments, the tau flavor 
violation might be found in the signal. The cross sections for $e^+e^-
(\mu^+\mu^-)\rightarrow \tilde{l}^+\tilde{l}^-\rightarrow 
\tau^\pm\mu^\mp (\tau^\pm e^\mp)+X$ are suppressed by at most the 
mass difference over the widths for the sleptons
\cite{Arkani-Hamed:1996au}. The search for these processes in the
collider experiments has more sensitivity for a small $\tan\beta$
region compared with the search for the LFV tau decay
\cite{Hisano:1998wn}.}

Finally, we discuss other tau LFV processes in the minimal SUSY seesaw
models.  In a broader parameter space, $\tau\rightarrow \mu(e)
\gamma$ is the largest tau LFV processes, unless it is
suppressed by some accidental cancellation or much heavier
SUSY particle masses. 
The LFV tau decay modes to three leptons are
dominantly induced by the photon-penguin contributions, and
they are correlated with $\tau\rightarrow\mu(e)\gamma$
as 
\begin{eqnarray}
  Br(\tau \rightarrow \mu2 e)/
  Br(\tau \rightarrow \mu\gamma) 
  &\simeq&
  1/94,
  \\ 
  Br(\tau \rightarrow 3 \mu)/
  Br(\tau \rightarrow \mu \gamma) 
  &\simeq&
  1/440,
  \\
  Br(\tau \rightarrow 3 e)/
  Br(\tau \rightarrow e \gamma)
  &\simeq&
  1/94,
  \\ 
  Br(\tau \rightarrow e2\mu)/
  Br(\tau \rightarrow e \gamma) 
  &\simeq&
  1/440.
\end{eqnarray}
The LFV tau decay modes into pseudo-scalar mesons tend to be 
smaller than those to three leptons since the branching
ratios are not proportional to $\tan^2\beta$.
 
When sleptons are much heavier than the weak scale,
$Br(\tau\rightarrow\mu(e)\gamma)$ is suppressed. In this case,
$\tau\rightarrow \mu(e)2\mu$ and $\tau\rightarrow \mu
(e) \eta$ induced by Higgs boson exchange become relatively
important \cite{Babu:2002et}. The LFV anomalous Yukawa coupling for the Higgs
bosons is generated by the radiative correction, and it is not
suppressed by powers of the slepton masses. While these processes are
suppressed by a small Yukawa coupling constant for muon or strange
quark, they may have sizable branching ratios when $\tan\beta$ is
large since the branching ratios are proportional to
$\tan^6\beta$. When $\delta_{\tau\mu}^{L}$ is non-vanishing, the
approximate formula for $Br(\tau\rightarrow 3\mu)$ is
given as 
\begin{eqnarray}
\lefteqn{Br(\tau\rightarrow 3\mu)
=
\frac{m_\mu^2 m_\tau^2 \epsilon_2^2 |\delta_{\tau\mu}^{L}|^2}
     {8\cos^6\beta}
Br(\tau\rightarrow\mu\nu_\tau\bar{\nu}_{\mu})}
\nonumber\\
\lefteqn{\times
\left[
\left(
\frac{\sin(\alpha-\beta) \cos\alpha}{M_{H^0}^2}
-\frac{\cos(\alpha-\beta) \sin\alpha}{M_{h^0}^2}
\right)^2
\right.}
\nonumber\\
\lefteqn{\left.
~~~~~~~~+
\frac{\sin^2\beta}{M_{A^0}^4}
\right]}\nonumber\\
\lefteqn{\simeq 3.8\times 10^{-7} \times |\delta_{\tau\mu}^{L}|^2
}
\nonumber\\
\lefteqn{
\times\left(\frac{\tan\beta}{60}\right)^6
\left(\frac{M_{A^0}}{100{\rm GeV}}\right)^{-4},}
\label{tauetamu}
\end{eqnarray}
and $Br(\tau\rightarrow\mu\eta)$ is roughly five times larger than
$Br(\tau\rightarrow 3\mu)$ though it depends on the anomalous coupling
of the Higgs bosons to the bottom and strange quark
\cite{Brignole:2004ah}.  Here, $\epsilon_2$ is a function of the SUSY
particle masses. We take a limit of large $\tan\beta$ and equal
SUSY-breaking mass parameters in the last step in
Eq.~(\ref{tauetamu}). Notice that $\tau\rightarrow\mu\gamma$ also has
a comparable branching ratio to them even if the Higgs mediation is
dominated, since the Higgs loop diagram is enhanced by the tau Yukawa
coupling constant.

\section{SUSY SU(5) GUT with Right-Handed Neutrinos}
Let us extend the minimal SUSY seesaw model to the SUSY SU(5) GUT. In
this model, doublet leptons and right-handed down-type quarks are
embedded in common {\bf 5}-dimensional multiplets, while doublet
quarks, right-handed up-type quarks, and right-handed charged leptons
are in the {\bf 10}-dimensional ones. Thus, the neutrino Yukawa
coupling induces the right-handed down-type squark mixing, and this
leads to rich flavor and CP violating phenomena in hadrons.

In this paper the minimal structure in the Yukawa coupling 
is assumed for
simplicity. The Yukawa coupling for quarks and leptons and the
Majorana mass term  for the right-handed neutrinos in this model are given as
\begin{eqnarray}
W_{\rm SU(5)}&=& 
\frac14 f_{ij}^{u} \Psi_i \Psi_j H 
+\sqrt{2} f_{ij}^{d} \Psi_i \Phi_j \overline{H}
\nonumber\\
&&
+f_{ij}^{\nu} \Phi_i \overline{N}_j {H}
+\frac12 M_{ij} \overline{N}_i \overline{N}_j,
\label{superp_gut}
\end{eqnarray}
where $\Psi$ and $\Phi$ are {\bf 10}- and {$\bf \bar{5}$}-dimensional
multiplets, respectively, and  $H$ ($\overline{H}$) is a {\bf 5}- ({$\bf
\bar{5}$}-) dimensional Higgs multiplet.

After removing the unphysical phases, the Yukawa coupling
constants and Majorana masses in Eq.~(\ref{superp_gut}) are given as
follows,
\begin{eqnarray}
f^u_{ij} &=& 
V_{ki} f_{u_k} {\rm e}^{i \varphi_{u_k}}V_{kj}, \\
f^d_{ij} &=& f_{d_i} \delta_{ij},\\  
f^\nu_{ij} &=& {\rm e}^{i \varphi_{d_i}} X^\star_{ik} f_{\nu_k} 
{\rm e}^{-i \varphi_{\nu_k}} W^\star_{kj} 
{\rm e}^{-i \overline{\varphi}_{\nu_k}} ,
\\
M_{ij} &=&   M_{N_i} \delta_{ij},  
\end{eqnarray}
where $\sum_i \varphi_{f_i} =0$ $(f=u,d,\nu)$ and $\sum_i
\overline{\varphi}_{f_i} =0$. Each unitary matrices $X$, $V$, and $W$
have only a phase, again. Here, $\varphi_{u}$ and $\varphi_{d}$ are
CP-violating phases inherent in the SUSY SU(5) GUT. The unitary matrix
$V$ is the CKM matrix in the extension of the SM to the SUSY SU(5)
GUT.

The colored-Higgs multiplets $H_c$ and $\overline{H}_c$ are introduced
in $H$ and $\overline{H}$ as SU(5) partners of the Higgs doublets in
the SUSY SM, respectively\footnote{
While the proton decay by the dimension-five operator, which is
induced by the colored-Higgs exchange, is a serious problem in the
minimal SUSY SU(5) GUT \cite{dim5}, it depends on the structure of the
Higgs sector \cite{dim5sup}. In this article, we ignore the proton decay
while we adopt the minimal structure of the Higgs sector.
}, and they have new flavor-violating interactions.
Eq.~(\ref{superp_gut}) is represent by the fields in the SUSY SM as
follows,
\begin{eqnarray}
\lefteqn{W_{\rm SU(5)}= W_{{\rm SUSY SM}+\overline{N}} 
+ \frac12 
V_{ki} f_{u_k} {\rm e}^{i \varphi_{u_k}}V_{kj} 
 Q_i Q_j H_c }
\nonumber\\
\lefteqn{+  
f_{u_i} V_{ij} {\rm e}^{i \varphi_{d_j}}
\overline{U}_i \overline{E}_j H_c 
+ 
f_{d_i} {\rm e}^{-i \varphi_{d_i}}
Q_i L_i \overline{H}_c }
\nonumber\\
\lefteqn{
+ 
{\rm e}^{-i \varphi_{u_i}}
V_{ij}^\star
f_{d_j} \overline{U}_i \overline{D}_j \overline{H}_c 
+ 
{\rm e}^{i \varphi_{d_i}}
X_{ij}^\star f_{\nu_j}
\overline{D}_i \overline{N}_j H_c.}
\nonumber\\
\end{eqnarray}
Here, the superpotential in the SUSY SM with the right-handed neutrinos
is
\begin{eqnarray}
W_{{\rm SUSY~SM}+\overline{N}} &=&
V_{ji} f_{u_j}  Q_i \overline{U}_j H_f 
+
f_{d_i} Q_i \overline{D}_i \overline{H}_f
\nonumber\\
&&
+
W_{\rm seesaw}
\label{superp_mssm}
\end{eqnarray}
The flavor-violating interactions absent in the SUSY SM emerge in the
SUSY SU(5) GUT due to existence of the colored-Higgs multiplets.

If the SUSY-breaking terms in the SUSY SM are generated by
interactions above the colored-Higgs mass, such as in the
supergravity, the sfermion mass terms may get sizable corrections by
the colored-Higgs interactions. Here we assume the minimal
supergravity scenario again. In this case, the flavor-violating
SUSY-breaking mass terms at low energy are induced by the radiative
correction, and they are approximately given as
\begin{eqnarray}
(\delta m_{{Q}}^2)_{ij}  &\simeq&-\frac{2}{(4\pi)^2} 
(3m_0^2+ A_0^2) 
\nonumber\\&&
V_{ki}^\star
f_{u_k}^2
V_{kj} 
\nonumber\\&&
(3 \log\frac{M_G}{M_{GUT}}
+ \log\frac{M_{GUT}}{M_{SUSY}}),\nonumber\\
(\delta m_{\overline{U}}^2)_{ij}  &\simeq&-\frac{4}{(4\pi)^2} 
(3m_0^2+ A_0^2)
\nonumber\\&&
{\rm e}^{i\varphi_{u_i}}V_{ik}
f_{d_k}^2
V_{jk}^\star {\rm e}^{-i\varphi_{u_j}} 
\nonumber\\&&
\log\frac{M_G}{M_{GUT}}, \nonumber\\
(\delta m_{\overline{D}}^2)_{ij}  &\simeq&-\frac{2}{(4\pi)^2} 
(3m_0^2+ A_0^2)
\nonumber\\&&
{\rm e}^{-i\varphi_{d_i}}X_{ik}
f_{\nu_k}^2
X^\star_{jk} {\rm e}^{i\varphi_{d_j}} 
\nonumber\\&&
\log\frac{M_G}{M_{GUT}},\nonumber\\
(\delta m_{L}^2)_{ij}  &\simeq&-\frac{2}{(4\pi)^2} 
(3m_0^2+A_0^2) 
\nonumber\\&&
X_{ik}
f_{\nu_k} {\rm e}^{i\varphi_{\nu_k}}
W_{kl}
\nonumber\\&&
W^\star_{ml}
{\rm e}^{-i\varphi_{\nu_m}} f_{\nu_m} 
X_{jm}^\star
\nonumber\\&&
\log\frac{M_G}{M_{N_l}},\nonumber\\
(\delta m_{\overline{E}}^2)_{ij}  &\simeq&-\frac{6}{(4\pi)^2} 
(3m_0^2+ A_0^2)
\nonumber\\&&
{\rm e}^{-i\varphi_{d_i}} V^\star_{ki}
f_{u_k}^2
V_{kj} {\rm e}^{i\varphi_{d_j}} 
\nonumber\\&&
\log\frac{M_G}{M_{GUT}},
\label{sfermionmass}
\end{eqnarray}
where $i\ne j$. Here, $M_{GUT}$ and $M_{G}$ are the GUT scale and
the reduced Planck scale, respectively. In the SUSY SM with
the right-handed neutrinos, the flavor-violating structures appear
only in the left-handed squark and left-handed slepton mass
matrices. On the other hand, in the SUSY SU(5) GUT, other sfermions
may also have sizable flavor violation. In particular, the CP-violating
phases inherent in the SUSY SU(5) GUT appear in
$(m_{\overline{U}}^2)$, $(m_{\overline{D}}^2)$, and
$(m_{\overline{E}}^2)$ \cite{Moroi:2000mr}.

In this section we assume the minimal structure for the Yukawa
coupling constants given in Eq.~(\ref{superp_gut}). The $b/\tau$ mass
ratio may be explained by it while the down-type quark and charged
lepton masses in the first and second generations are not
compatible. The modification of the Yukawa sector, such as
introduction of the higher-dimensional operators, the vector-like
matters, or the complicate Higgs structure, may change the low-energy
prediction for the flavor violation, especially, between the first and
second generations.  Thus, we will concentrate on the transition between
the second and third generations in the following. We assume that the
Yukawa coupling constants, including the neutrino one, are
hierarchical, and that the extra interaction gives negligible
contribution to the transition between the second and third
generations.

The neutrino-induced off-diagonal terms for the sfermion masses are
$(m_{\overline{D}}^2)_{ij}$ and $(m_{L}^2)_{ij}$ ($i\ne j $).  Let us
demonstrate the good correlation between $(m_{\overline{D}}^2)_{23}$
and $(m_{L}^2)_{23}$. The non-trivial structure in the right-handed
neutrino mass matrix may dilute the correlation as in
Eq.~(\ref{sfermionmass}).  However, if the right-handed neutrino masses are
hierarchical, the correlation is expected to be good, as will be shown.

We use two-generation model for simplicity \cite{Ellis:2001xt},
ignoring the first generation. Here, we adopt the parameterization for
the neutrino sector by Casas and Ibarra \cite{ci}.  In this
parametrization the neutrino sector can be parametrized by the
left-handed ($m_{\nu_i}$) and right-handed neutrino masses ($M_{N_i}$),
the MNS matrix with the Majorana phases ($Z=UP$), and a complex orthogonal
matrix ($R$) as
\begin{eqnarray}
f_{ij}^{\nu}
&=&
\frac{1}{\langle H_f\rangle}
Z_{ik}^\star \sqrt{m_{\nu_k}}R_{kj}\sqrt{M_{N_j}}.
\label{casas}
\end{eqnarray}
Using this formula, $(\delta m_{\overline{D}}^2)_{23}$ and $(\delta
m_{{L}}^2)_{23}$ are given as
\begin{eqnarray}
\lefteqn{(\delta m_{\overline{D}}^2)_{23} 
=
\frac{1}{2(4\pi)^2}
{\rm e}^{-i(\varphi_{d_2}-\varphi_{d_3})}
\frac{(3m_0^2+A_0^2)}{\langle H_f\rangle^2} }
\nonumber\\
\lefteqn{\log\frac{M_G}{M_{GUT}}
\left((m_{\nu_\mu}+m_{\nu_\tau})(M_{N_2}-M_{N_3}) \cos 2\theta_r\right.}
\nonumber\\
\lefteqn{
\left.+(m_{\nu_\mu}-m_{\nu_\tau})(M_{N_2}+M_{N_3}) \cosh 2\theta_i\right.}
\nonumber\\
\lefteqn{
-2i\sqrt{m_{\nu_\mu}m_{\nu_\tau}}
\left.
((M_{N_2}-M_{N_3}) \sin\phi \sin 2\theta_r
\right.}
\nonumber\\
\lefteqn{
\left.
-(M_{N_2}+M_{N_3}) \cos\phi \sinh 2\theta_i\right),}
\label{md}
\\
\lefteqn{(\delta m_{{L}}^2)_{23} 
=
\frac{1}{2(4\pi)^2}
\frac{(3m_0^2+A_0^2)}{\langle H_f\rangle^2} 
}
\nonumber\\
\lefteqn{
\left((m_{\nu_\mu}+m_{\nu_\tau})(\overline{M}_{N_2}-\overline{M}_{N_3}) \cos 2\theta_r
\right.}
\nonumber\\
\lefteqn{
\left.
+(m_{\nu_\mu}-m_{\nu_\tau})(\overline{M}_{N_2}+\overline{M}_{N_3}) \cosh 2\theta_i\right.}
\nonumber\\
\lefteqn{
\left.
-2i\sqrt{m_{\nu_\mu}m_{\nu_\tau}}
\right.
((\overline{M}_2-\overline{M}_3) \sin\phi \sin 2\theta_r}
\nonumber\\
\lefteqn{
\left.
-(\overline{M}_2+\overline{M}_3) \cos\phi \sinh 2\theta_i\right)}
\label{ml}
\end{eqnarray}
with $\overline{M}_{N_i}=M_{N_i} \log M_G/M_{N_i}$. Here, we use
\begin{eqnarray}
R
&=&
\left(
\begin{array}{cc}
\cos(\theta_r+i \theta_i)
&\sin(\theta_r+i \theta_i)\\
-\sin(\theta_r+i \theta_i)
&\cos(\theta_r+i \theta_i)
\end{array}
\right),
\\
X&=&
\left(
\begin{array}{cc}
1/\sqrt{2}&1/\sqrt{2}\\
-1/\sqrt{2}&1/\sqrt{2}
\end{array}
\right)
\left(
\begin{array}{cc}
{\rm e}^{\phi_M}&\\
&1
\end{array}
\right),
\end{eqnarray}
assuming a maximal mixing for the atmospheric neutrino. $\phi_M$
is the Majorana phase for the light neutrinos. If the right-handed neutrino
masses are hierarchical $(M_{N_3}\gg M_{N_2})$, the correlation is good as 
\cite{Hisano:2003bd}
\begin{eqnarray}
\frac{(\delta m_{\overline{D}}^2)_{23}}{(\delta m_{{L}}^2)_{23}} 
&\simeq&
{\rm e}^{-i(\varphi_{d_2}-\varphi_{d_3})}
\frac{\log\frac{M_G}{M_{GUT}}}{\log\frac{M_G}{M_{N_3}}}.
\end{eqnarray}
Thus, it is important to check this GUT relation in the lepton and
hadron flavor physics for confirmation of the SUSY GUT with the
right-handed neutrinos.

\section{Mercury EDM in the SUSY GUTs}

The Belle experiment in the KEK $B$ factory reported recently that the
CP asymmetry in $B_d\rightarrow \phi K_s$ $(S_{\phi K_s})$ is $-0.96\pm
0.50^{+0.09}_{-0.11}$, and $3.5 \sigma$ deviation from the
Standard-Model (SM) prediction $0.731\pm0.056$ is found
\cite{Abe:2003yt}. The CP violation in $B_d\rightarrow \phi K_s$ is sensitive 
to the new physics since $b\rightarrow s\bar{s}s$ is a radiative process
\cite{gw}. In fact, the SUSY models may predict a sizable deviation 
of the CP violation in $B_d\rightarrow \phi K_s$ from the SM
prediction. If the right-handed bottom and strange squarks have a
sizable mixing in the SUSY GUTs with the right-handed neutrinos, the
gluon-penguin diagram may give a non-negligible contribution to
$b\rightarrow s\bar{s}s$ in a broad parameter space where the
contribution to $b\rightarrow s\gamma$ is a sub-dominant.  Nowadays,
$B_d\rightarrow\phi K_s$ in the SUSY models is studied extensively.

However, the correlation between the CP asymmetry in $B_d\rightarrow
\phi K_s$ ($S_{\phi K_s}$) and the chromoelectric dipole moment (CEDM)
of strange quark ($d_s^C$) is strong in the SUSY models with the
right-handed squark mixing \cite{Hisano:2003iw}. In typical SUSY
models, the left-handed squarks also have flavor mixing due to the
top-quark Yukawa coupling and the CKM mixing (see
Eq.~(\ref{sfermionmass})), and the left-handed bottom and strange
squark mixing is as large as $\lambda^2\sim 0.04$.  When both the
right-handed and left-handed squark mixings between the second and
third generations are non-vanishing, the CEDM of the strange quark is
generated. Since $S_{\phi K_s}$ and $d_s^C$ may have a strong
correlation in the SUSY models with the right-handed squark mixing,
the constraint on $d_s^C$ by the measurement of the EDM of $^{199}$Hg
limits the gluon-penguin contribution from the right-handed squark
mixing to $S_{\phi K_s}$
\cite{Hisano:2003iw}.

Before deriving the constraint on the bottom and strange squark
mixing, we discuss the EDM of the nuclei. The EDMs of the diamagnetic
atoms, such as $^{199}$Hg, come from the CP-violating nuclear force by
pion or eta meson exchange. The quark CEDMs,
\begin{eqnarray}
 H_{\rm eff}=  \sum_{q=u,d,s} d_q^C \frac{i}{2}g_s\overline{q}\sigma^{\mu\nu}T^A\gamma_5 q G^A_{\mu\nu},
\end{eqnarray}
generate the CP-violating meson-nucleon  coupling, and the EDM of $^{199}$Hg
is evaluated in Ref.~\cite{Falk:1999tm} as 
\begin{eqnarray}
d_{\rm Hg}=-3.2\times 10^{-2} e\times (d_d^C-d_u^C-0.012 d_s^C).
\label{deg}
\end{eqnarray}
The chiral perturbation theory implies that $\bar{s}s$ in the matrix
element of nucleon is not suppressed, and it leads to non-vanishing
contribution from the CEDM of the strange quark. The suppression
factor in front of $d_s^C$ in Eq.~(\ref{deg}) comes from the eta meson
mass and the CP-conserving coupling of the eta meson and nucleon. From
current experimental bound on $d_{\rm Hg}$ ($d_{\rm Hg}<2.1\times
10^{-28}e~cm)$ \cite{Romalis:2000mg}
\begin{eqnarray}
e|d_d^C-d_u^C-0.012 d_s^C|&<&7\times 10^{-27}{e~cm}.
\end{eqnarray}
If $d_d^C$ and $d_u^C$ are negligible in the equation, 
\begin{eqnarray}
e|d_s^C|&<&7\times 5.8\times 10^{-25}{e~cm}.
\label{scedm}
\end{eqnarray}

The neutron EDM should also suffer from the CEDM of the strange
quark. However, it is argued in Ref.~\cite{Pospelov:2000bw} that the
Peccei-Quinn symmetry suppresses it. In the paper the QCD sum rule is
adopted to evaluate the neutron EDM. Thus, it does not include the see
quark contribution to the neutron EDM, and it is expected that the
Peccei-Quinn symmetry decouples the CEDM of the strange quark from the
neutron EDM in this evaluation. There is no reliable calculation for
the see quark contribution at present. If the ``standard'' loop
calculation of the neutron EDM in the chiral Lagrangian is reliable,
the current experimental bound on the neutron EDM may give a
comparable constraint on the CEDM of the strange quark.

\begin{figure}[htb]
\begin{center}
\includegraphics[width=15pc]{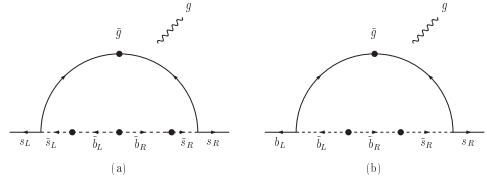} 
\caption{a) The dominant diagram contributing to the CEDM of the strange
quark when both the left-handed and right-handed squarks have mixings.
b) The dominant SUSY diagram contributing to the CP asymmetry in
$B_d\rightarrow \phi K_s$ when the right-handed squarks have mixing. }
\label{fig:edm1}
\end{center}
\end{figure}

In the SUSY models, when the left-handed and right-handed squarks have
mixings between the second and third generations, the CEDM of the strange
quark is generated by a diagram in Fig.~\ref{fig:edm1}(a), and it is enhanced by
$m_b/m_s$. Using the mass insertion technique, $d_s^C$ is given 
up to the QCD correction as
\begin{eqnarray}
  e d_s^C &\simeq&
e  \frac{\alpha_s}{4\pi} \frac{m_{\tilde{g}}}{m^2_{\tilde{q}}}
\left(-\frac{11}{30}\right)
\nonumber\\
&\times&{\mathrm{Im}}\left[( \delta_{LL}^{(d)})_{23}
\,(\delta_{LR}^{(d)})_{33}\,(\delta_{RR}^{(d)})_{32}\right]
\label{dsap}
\\
&=& -4.0 \times 10^{-23} \sin\theta\, {e\, {\rm cm}} \,
\left(\frac{m_{\tilde{q}}}{500\mathrm{GeV}}\right)^{-3}
\nonumber\\
&\times&
\left(\frac{(\delta_{LL}^{(d)})_{23}}{0.04}\right)
\left(\frac{(\delta_{RR}^{(d)})_{32}}{0.04}\right)
\left(\frac{\mu \tan\beta}{5000\mathrm{GeV}}\right)
\nonumber\\
\end{eqnarray}
in a limit of $m_{\tilde{q}}/m_{\tilde{g}}=1$.
The mass insertion parameters $(\delta_{LL}^{(d)})_{23}$,
$(\delta_{RR}^{(d)})_{32}$, and $( \delta_{LR}^{(d)})_{33}$ are 
\begin{eqnarray}
\lefteqn{(\delta_{LL}^{(d)})_{23}=
\frac{\left(m_{\tilde{d}_L}^2\right)_{23}}{m^2_{\tilde{q}}},\,\,
(\delta_{RR}^{(d)})_{32}=
\frac{\left(m_{\tilde{d}_R}^2\right)_{32}}{m^2_{\tilde{q}}},}
\nonumber\\
\lefteqn{( \delta_{LR}^{(d)})_{33} 
= \frac{m_b\left(A_b -\mu\tan\beta\right)}{m^2_{\tilde{q}}},}
\end{eqnarray}
and $\theta={\mathrm{arg}}[( \delta_{LL}^{(d)})_{23}
\,(\delta_{LR}^{(d)})_{33}\,(\delta_{RR}^{(d)})_{32}]$.  
In the typical SUSY models, $(\delta_{LL}^{(d)})_{23}$ is
$O(\lambda^2)\simeq 0.04$. From this formula, it is obvious that the
right-handed squark mixing or the CP-violating phase should be
suppressed. For example, for $m_{\tilde{q}}=500$GeV, $\mu
\tan\beta=5000$GeV, and ${(\delta_{LL}^{(d)})_{23}}={0.04}$,
\begin{eqnarray}
|\sin\theta (\delta_{RR}^{(d)})_{32}| <5.8\times 10^{-4}.
\end{eqnarray}

In the SUSY SU(5) GUT with the right handed neutrinos, the tau
neutrino Yukawa coupling induces the right-handed down-type squark
mixing between the second and third generations radiatively as 
\begin{eqnarray}
(\delta m_{\overline{D}}^2)_{23}  &\simeq&-\frac{2}{(4\pi)^2} 
{\rm e}^{-i(\varphi_{d_2}-\varphi_{d_3})} X_{32} X^\star_{33}
\nonumber\\
\lefteqn{
\times\frac{m_{\nu_\tau} M_N}{\langle H_f \rangle^2}
 (3m_0^2+A_0^2) \log\frac{M_G}{M_{GUT}},}
\label{msd}
\end{eqnarray}
when the hierarchical neutrino Yukawa coupling and $U\simeq X$  are
assumed.  This correction depends on the GUT generic phase. From this equation,
the CEDM of the strange quark is larger than the experimental bound
when $M_{N_\tau}$ is larger than about $10^{(12-13)}$ GeV and
$(\varphi_{d_2}-\varphi_{d_3})$ is of the order of 1.  This means that
the measurement of the EDM of $^{199}$Hg atom is very sensitive to the
right-handed neutrino sector in the SUSY SU(5) GUT.

The current experimental bound on the EDM of $^{199}$Hg atom is
determined by the statistics, and the further improvement is expected
\cite{Romalis:2000mg}. Also, it is argued  recently  in
Ref.~\cite{Semertzidis:2003iq} that the measurement of the
deuteron EDM may improve the bound on the CP-violating nuclear force
by two order of magnitude. If it is realized, it will be a stringent
test on the SUSY models with the right-handed squark mixing, such as
the SUSY GUTs.

Now let us discuss the correlation between $d_s^C$ and $S_{\phi
K_s}$ in the SUSY models with the right-handed squark mixing. The
right-handed bottom and strange squark mixing may lead to the sizable
deviation of $S_{\phi K_s}$ from the SM prediction by the
gluon-penguin diagram, especially for large $\tan\beta$.  The box
diagrams with the right-handed squark mixing also contribute to
$S_{\phi K_s}$, however, they tend to be sub-dominant and do not
derive the large deviation of $S_{\phi K_s}$ from the SM prediction.
Thus, we neglect the box contribution in this article for simplicity.

\begin{figure}[htb]
\begin{center}
\includegraphics[width=15pc]{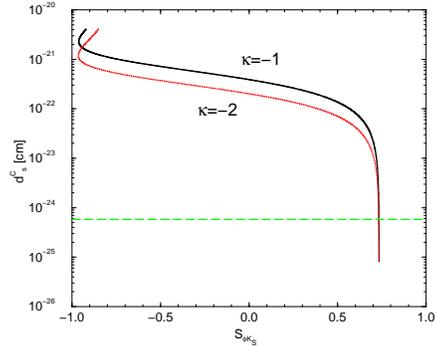} 
\caption{The correlation between $d_s^C$ and $S_{\phi K_s}$ assuming
$d_s^C = -{m_b}/(4\pi^2)
{\mathrm{Im}}[ ( \delta_{LL}^{(d)})_{23}C_8^{R}]$. Here, $( \delta_{LL}^{(d)})_{23}
=-0.04$ and ${\mathrm{arg}}[C_8^{R}]=\pi/2$. $\kappa$ comes from 
the matrix element of chromomagnetic
moment  in $B_d\rightarrow \phi K_s$.
The dashed line is the upperbound on $d_s^C$ from the EDM of 
$^{199}$Hg atom. }
\label{fig:edm2}
\end{center}
\end{figure}

The effective operator, which induces the gluon-penguin diagram by the
right-handed squark mixing, is
\begin{eqnarray}
 H_{\rm eff}&=& - C_8^{R} \frac{g_s}{8\pi^2}
    m_b\overline{s}\sigma^{\mu\nu}T^A P_L b G^A_{\mu\nu}.
\end{eqnarray}
When the right-handed squarks have the mixing, the dominant
contribution to $C_8^{R}$ is supplied by a diagram with the double
mass insertion of $(\delta_{RR}^{(d)})_{32}$ and
$(\delta_{RL}^{(d)})_{33}$ (Fig.~\ref{fig:edm2}(b)). Especially, it is significant
when $\mu\tan\beta$ is large. The contribution of Fig.~\ref{fig:edm1}(b) to $C_8^{R}$ 
is  given up to the QCD correction as
\begin{eqnarray}
 C_8^{R}&=&\frac{7\pi \alpha_s}{30{m_b} m_{\tilde{q}}}
(\delta_{LR}^{(d)})_{33}
(\delta_{RR}^{(d)})_{32}.
\label{c8ap}
\end{eqnarray}
in a limit of $m_{\tilde{q}}/m_{\tilde{g}}=1$.  Comparing
Eq.~(\ref{dsap}) and Eq.~(\ref{c8ap}), a strong correlation between
$d_s^C$ and $C_8^R$ is derived as
\begin{eqnarray}
d_s^C &=& -\frac{m_b}{4\pi^2} \frac{11}{7}
{\mathrm{Im}}\left[( \delta_{LL}^{(d)})_{23} C_8^{R}\right]
\label{massin}
\end{eqnarray}
up to the QCD correction \cite{Hisano:2003iw}. The coefficient $11/7$
in Eq.~(\ref{massin}) changes from 3 to 1 for $0<x<\infty$.

In Fig.~\ref{fig:edm2}, the correlation between $d_s^C$ and $S_{\phi
K_s}$ is presented. Here $d_s^C = -{m_b}/(4\pi^2)
{\mathrm{Im}}[(\delta_{LL}^{(d)})_{23}C_8^{R}]$ is assumed up to the
QCD correction.  Here, we take $(\delta_{LL}^{(d)})_{23} =-0.04$,
${\mathrm{arg}}[C_8^{R}]=\pi/2$ and $|C_8^R|$ corresponding to
$10^{-5}<|(\delta_{RR}^{(d)})_{32}|<0.5$. $\kappa$ comes from the
matrix element of chromomagnetic moment in $B_d\rightarrow \phi K_s$,
and $\kappa=-1.1$ in the heavy-quark effective theory \cite{hlmp}.
Since $\kappa$ may suffer from the large hadron uncertainty, we take
$\kappa=-1$ and $-2$.  From this figure, it is found that the
deviation of $S_{\phi K_s}$ from the SM prediction due to the gluon-penguin contribution should be tiny when the constraint on $d_s^C$ in
Eq.~(\ref{scedm}) is applied.

Finally, we discuss the $\tau\rightarrow \mu \gamma$ in the SUSY SU(5)
GUT with the right-handed neutrinos. In the last section we show the
strong correlation between ${(\delta m_{\overline{D}}^2)_{23}}$ and
${(\delta m_{{L}}^2)_{23}}$ up to the GUT generic phases in this
model.  Thus, the constraint on the strange quark CEDM gives a bound
on the prediction of $Br(\tau\rightarrow\mu\gamma)$ smaller than the
future sensitivity in the super $B$ factories when the GUT generic
phase is of the order of one.

\section{Summary}

In this article we reviewed the third generation flavor violation in
the SUSY seesaw model and SUSY SU(5) GUT with the right-handed
neutrinos. After discovering neutrino oscillation, the flavor
violation induced the neutrino Yukawa coupling becomes important. Now
the $B$ factories starts to access interesting region for
$\tau\rightarrow \mu(e)\gamma$. It may give a new information about
structure of the seesaw mechanism. Also, we show new stringent
constraint from the Mercury EDM on the SUSY models with the
right-handed squark mixing, such as the SUSY SU(5) GUT with the
right-handed neutrinos, since the Mercury EDM is sensitive to the
strange quark CEDM. This constraint implies that the deviation of
$S_{\phi K_s}$ from the SM prediction, which is observed in the Belle
experiment, should be suppressed in the SUSY GUT with right-handed
neutrinos.  Sensitivity to the CP-violating nuclear force may be
improved furthermore by about two orders in the measurement of
deuterium EDM, and this may be a big impact on SUSY GUT.

\end{document}